\documentclass[a4paper,11pt]{article}
\pdfoutput=1 

\usepackage{jheppub} 

\usepackage[T1]{fontenc} 

\usepackage[all]{xy}

\def\be{\begin{equation}}
\def\ee{\end{equation}}
\def\ba{\begin{eqnarray}}
\def\ea{\end{eqnarray}}

\def\rd{\mathrm{d}}
\def\rD{{\rm D}}
\def\CL{{\cal L}}
\def\g{\mathfrak{g}}
\def\Lieg{{\mathrm{Lie}(G)}}

\newcommand{\md}{\mathrm{d}}
\newcommand{\TTM}{\mathbb{T}M}

\let\a=\alpha \let\b=\beta

  \let\S=\Sigma
\let\r=\rho

\newcommand*{\R}{{\mathbb R}}

\title{\boldmath 2d Gauge Theories and Generalized Geometry}

\author[a]{Alexei Kotov}
\author[b]{Vladimir Salnikov}
\author[c]{Thomas Strobl}

\affiliation[a]{Department of Mathematics and Statistics, Faculty of Science and Technology \\ University of Troms{\o}, N-9037 Troms{\o}, Norway}
\affiliation[b]{Laboratoire de Math\'ematiques Nicolas Oresme \\
Universit\'e de Caen Basse-Normandie, \\
CS 14032, Bd. Mar\'echal Juin,  BP 5186,
  14032 Caen Cedex,
  France
}
\affiliation[c]{Institut Camille Jordan,
Universit\'e Claude Bernard Lyon 1 \\
43 boulevard du 11 novembre 1918, 69622 Villeurbanne cedex,
France}

\emailAdd{oleksii.kotov@uit.no}
\emailAdd{vladimir.salnikov@unicaen.fr}
\emailAdd{strobl@math.univ-lyon1.fr}

\abstract{We show that in the context of two-dimensional sigma models minimal coupling of an ordinary rigid symmetry Lie algebra $\g$  leads naturally to the appearance of the ``generalized tangent bundle'' $\mathbb{T}M \equiv TM \oplus T^*M$ by means of composite fields. Gauge transformations of the composite fields follow the Courant bracket, closing upon the choice of a Dirac structure $D \subset \mathbb{T}M$ (or, more generally, the choide of a ``small Dirac-Rinehart sheaf'' $\cal{D}$), in which the fields as well as the symmetry parameters are to take values. In these new variables, the gauge theory takes the form of a (non-topological) Dirac sigma model, which is applicable in a  more general context and proves to be universal in two space-time dimensions: A gauging of $\g$ of a standard sigma model with Wess-Zumino term exists, \emph{iff} there is a prolongation of the rigid symmetry to a Lie algebroid morphism from the action Lie algebroid $M \times \g\to M$ into $D\to M$ (or the algebraic analogue of the morphism in the case of $\cal{D}$). The gauged sigma model results from a pullback by this morphism from the Dirac sigma model, which proves to be universal in two-spacetime dimensions in this sense.

 \vspace{5mm}
 }

\begin{document}

\maketitle

\section{Introduction}
The Dirac sigma model (DSM) \cite{Kotov-Schaller-Strobl} was originally constructed so as to jointly generalize the  Poisson sigma model \cite{Schaller-Strobl} and the G/G Wess-Zumino-Novikov-Witten (WZNW) model \cite{Novikov,Witten}, both of which are
two-dimensional gauge theories with no physical degrees of freedom, so-called topological sigma models. As we will see, however, there exists also a non-topological version of this sigma model, e.g.~in generalization of the G/H WZNW model. In fact, this generalization goes much further than one may think: it encompasses \emph{all} gauged standard sigma models with Wess-Zumino  term in two dimensions. Here we consider at this stage only standard gauging, starting with a rigid symmetry group $G$ of an ungauged theory. In the presence of a Wess-Zumino term such a gauging does not always exist, but if it exists, the resulting functional is encompassed by the formalism. Non-standard gauge theories, as discovered recently in \cite{PRL2014}, will be addressed in their two-dimensional form and also related to the DSM in a separate work.

Standard gauging introduces Lie algebra valued 1-forms $A^a e_a \in \Omega^1(\Sigma,\g)$. The same Lie algebra is acting on the target space $M$ of the sigma model, $e_a \mapsto v_a \in \Gamma(TM)$. Composition thus provides a 1-form on the two-dimensional space-time $\Sigma$ with values in the vector fields over $M$, $V=v_a A^a$. More precisely, if the map $X \colon \Sigma \to M$ corresponds to the (scalar) fields from the ungauged theory, one obtains a gauge field $V \in \Omega^1(\Sigma, X^*TM)$. In $d=2$ it is near at hand to amend a standard sigma model, which needs a metric $g$ on the target $M$ for its definition, with (the $X$-pullback of) a 2-form $B$ on $M$ or, more generally and permitting Wess-Zumino terms, a closed 3-form $H \in \Omega_{cl}^3(M)$. Rigid invariance of $B$ or $H$ with respect to $\g$ implies the existence of a map from $\g$ into the 1-forms on $M$, $e_a \mapsto \alpha_a \in \Gamma(T^*M)$. Again, by composition we obtain a
gauge field $A = \alpha_a A^a \in \Omega^1(\Sigma, X^*T^*M)$, or, if we consider both of them as belonging together, a gauge field ${\cal A} = V \oplus A \in \Omega^1(\Sigma, X^*\mathbb{T}M)$.

Here $\mathbb{T}M = TM \oplus T^*M$ is the vector bundle sum of the tangent and the cotangent bundle, which sometimes is called the ``generalized tangent bundle''. Given an $H$ as above, it carries naturally the structure of a so-called ($H$-twisted) standard Courant algebroid, a notion that we will recall in the main text. Among others, this bundle is equipped with a canonical scalar product $\langle \cdot , \cdot \rangle$ and natural bracket $[ \cdot , \cdot ]$, the Courant-Dorfmann bracket, on its sections. We will show that these structures appear in the gauge transformations of ${\cal A}$ as induced from those of the standard gauge fields.

The fields $A$ and $V$ are not indendent from one another. In particular, they turn out to always take values in a Dirac structure.\footnote{This picture is ``essentially correct'', but needs some refinement, cf.~in particular  Appendix A.} This is a subbundle of $TM \oplus T^*M$
which is maximally isotropic with respect to $\langle \cdot , \cdot \rangle$ and closed with respect to the bracket, $[ \Gamma(D) , \Gamma(D) ]\subset \Gamma(D)$. Moreover, in terms of the new variables---$X$ together with the composite field ${\cal A}$---the gauged sigma model takes somewhat surprisingly precisely the form of the Dirac sigma model. If the fields $V \oplus A$ sweep out all of $D$, as considered in the original work \cite{Kotov-Schaller-Strobl} on the DSM, the model is topological. However, in general, it is not necessary to require this, which then permits the model to carry physical degrees of freedom.

The present paper is in some sense the Lagrangian counterpart of the paper \cite{Alekseev-Strobl} relating current algebras of two-dimensional sigma models in their Hamiltonian formulation to Courant algebroids. In particular, the Poisson bracket of symmetry generating currents $J_{\psi}[\varphi]$ of a variety of two-dimensional models reflects a similar behavior as the one mentioned above. More precisely, it was found that they follow the algebra
\begin{equation}\label{currents}
\{ J_{\psi}[\varphi], J_{\bar{\psi}}[\bar{\varphi}]\} =  J_{[\psi,\bar{\psi}]}[\varphi \bar{\varphi} ] -F_{\langle \psi, \bar{\psi}\rangle}[\rd_{S^1}(\varphi)  \bar{\varphi}] \, ,
\end{equation}
where the scalar product and bracket in the twisted standard Courant algebroid are noted as before; $\psi$,$~\!\bar{\psi}$ are sections in $\mathbb{T}M$, $\varphi$,$~\!\bar{\varphi}$ test functions over $ S^1$, in combination they parametrize the currents $J$. $F$ on the r.h.s.~is easy to explain,  $F_f[\mu] \equiv \oint_{S^1} (X^*f) \, \mu$ (for any $f \in C^\infty(M)$ and $\mu \in \Omega^1(S^1)$).
The currents $J_\psi[\varphi]$ are slightly more intricate to define, but can be regarded upon as follows: $J_\psi[\varphi] = \oint_{S^1} \varphi X^* \langle \psi , {\cal P} \rangle$, where $ {\cal P} = d_{S^1}X \oplus p \in \Omega^1 (S^1, X^*\mathbb{T}M)$, with $p$ being the canonical momentum 1-form conjugate to the ``closed string'' $X \colon S^1 \to M$.

Clearly, if the functions (currents) $J_\psi[\varphi]$ are to generate some symmetries on the phase space $T^*LM$, where $LM = \{ X \colon S^1 \to M\}$ denotes the loop space, for some subset of sections $\psi \in \Gamma(\mathbb{T}M)$,  on the right-hand-side of Equation \eqref{currents} the anomalous $F$-terms must vanish and the $J$-terms have to be parametrized by a section in the chosen subset. This is in particular the case if one wants to use them for gauging, corresponding to (first class) constraints on the Hamiltonian level. Evidently in the case of maximality, this implies that necessarily the subset of admissible sections $\psi$ have to lie inside a Dirac structure $D \subset \mathbb{T}M$. In any case, they have to define an involutive and isotropic subset of sections in $\TTM$, the sheaf of which we call a Dirac-Rinehart sheaf (cf.~Appendix A). 

In the present paper instead of loops we have maps from an oriented 2-surface to the target manifold $M$, $X \colon \Sigma \to M$. Instead of the phase space variables, we will consider ${\cal A}_\Phi := \langle X^*\Phi , {\cal A} \rangle$ here, where $\Phi \in \Gamma(\mathbb{T}M)$, to study the gauge transformations. Clearly, knowing the transformation behavior of ${\cal A}_\Phi$, one can reconstruct the one of $X$ and the gauge fields ($V$, $A$, and $A^a$). Similar to the currents above, we will parametrize gauge transformations by sections $\psi$ in $\TTM$ and test functions $\varphi \in C^\infty(\Sigma)$. One then finds
\begin{equation}\label{deltaAP}
\delta_{(\psi,\varphi)}{\cal A}_\Phi =  X^*\langle \psi, \Phi \rangle\, \md_\Sigma \varphi  + {\cal A}_{[\psi,\Phi]}\, \varphi\,.
\end{equation}
This will be obtained in the special case of \emph{ordinary} gauge symmetries on the composite fields, in which case sections $\psi$ (as well as $V \oplus A$) take values in involutive, isotropic subspaces $\cal{D}$. But one can also postulate the transformations \eqref{deltaAP} more generally. Calculating the commutator of such gauge transformations in this more general setting, one finds
\begin{equation}\label{comm}
[\delta_{(\psi,\varphi)}, \delta_{(\bar{\psi},\bar{\varphi})}] \,{\cal A}_\Phi =  \delta_{([{\psi},\bar{\psi}],\varphi\bar{\varphi})}{\cal A}_\Phi -X^*\langle 0\oplus \md \langle \psi,\bar{\psi}\rangle, \Phi \rangle \, \md_\Sigma(\varphi) \bar{\varphi}\, .\end{equation}
Thus, if one wants the transformations \eqref{deltaAP} to close for any choice of test functions $\varphi$ and a subset of sections $\psi \in \Gamma(\TTM)$, one needs the sections to take values in some Dirac structure $D \subset \TTM$ (or at least inside some $\cal{D}$) by the same argument as above for the currents.

For those symmetries that are obtained from the gauging of a rigid symmetry, this is always satisfied: There exists a bracket preserving map $\mu$ from $M \times \g$ into that $D$ where the composite gauge field $\cal{A}$ takes values in. In more technical terms, this map $\mu$ is a Lie algebroid morphism (a Dirac structure turns out to carry the structure of a Lie algebroid). Any standard sigma model with Wess-Zumino term which has been gauged can be obtained from the DSM by the pullback of such a morphism $\mu$. Note also that this can include cases where $\dim \g > \dim M$. In the most extreme case, when one regards a limit where the metric of the sigma model is turned to vanish and only the Wess-Zumino term $H$ remains, this group can be even infinite dimensional, while its gauging by means of the composite fields leads to the finite number of gauge fields only. This was illustrated in detail in \cite{Salnikov-Strobl} also.

The Dirac sigma model as well as its special case the Poisson sigma model (PSM) are classical examples for gauge theories with so-called open algebras. This seems to contradict the role of the DSM as a universal model for any ordinary gauged sigma model, where the symmetries follow the structural Lie algebra $\g$ and are certainly closed. We will resolve this apparent paradox in the main part of the paper. For example, in the case of the PSM, the main lesson is that in contrast to the conventional use of elements in the pullback bundle $\varepsilon \in \Gamma(X^*T^*M)$ to parametrize the symmetries, which then corresponds to a collection of $\dim M$ functions $\varphi \in C^\infty(\Sigma)$, one should parametrize them as above, that is by elements in the tensor product $C^\infty(\Sigma) \otimes \Gamma(\TTM)$. It is the non-trivial dependence on the second factor that closes the algebra, in accordance with the previous observations \cite{Bojowald-Kotov-Strobl}.

The structure of the paper is as follows: We first recall the setting of rigid symmetries in the context of 2d standard sigma models with Wess-Zumino term. We then show that minimal coupling, applied to the simplest type of actions in this family where the Wess-Zumino term is local and strictly invariant, leads to the correct form of the Dirac sigma model upon introduction of the above-mentioned composite fields $V$ and $A$. In section 4 we recall some basic notions of the Courant algebroid $\TTM$ and its Dirac structures and show how to obtain any gauged sigma model with WZ-term from the DSM by means of a pullback. In section 5 we address the gauge symmetries, arriving at the formulas \eqref{deltaAP} and \eqref{comm} anticipated already above. In a final section we present our conclusions as well as an outlook on follow-up research. Appendix A contains a mathematical, technical refinement.

\section{Action Functionals with Rigid Symmetries}
The standard sigma model in defined as a functional of smooth maps $X \colon \Sigma \to M$ where $(\Sigma,h)$ is some oriented Lorentzian signature pseudo-Riemannian $d$-manifold and $(M,g)$ is a Riemannian or pseudo-Riemannian $n$-manifold. The action functional is then given by
 \be \label{eq:standardsigma}
S_g[X] = \int\limits_\Sigma  \frac{1}{2} g_{ij}(X) \, \md X^i \wedge * \md X^j
\equiv \int_\Sigma \partial_\mu X^i \partial_\nu X^j \,h^{\mu \nu} g_{ij}(X) \, \sqrt{|\det(h)|}\md^d \sigma .
\ee
In this paper we consider $d=2$. We then can extend the above action by adding the pullback of a 2-form $B$ on $M$ assumed to be part of the given data on the target:
 \be \label{eq:sigma}
S[X] = \int\limits_\Sigma  \frac{1}{2} g_{ij}(X) \, \md X^i \wedge * \md X^j
+  \int\limits_{\S} X^* B .
\ee
Let us suppose that the metric $g$ has a nontrivial isometry group $G$, which infinitesimally implies
\be \CL_v g = 0 \, , \label{eq:Kill}
\ee
valid for the vector fields $v=\r(\xi)$ on $M$ corresponding to arbitrary elements  $\xi\in \g = \Lieg$, $\r$ denoting the representation of $\g$ on $M$ induced by the $G$-action. If in addition $B$ is $G$-invariant  up to a ``total divergence'', i.e.~if there exist  $(d-1)$-forms $\b$ for any $v$ such that
\be \CL_v B = \rd \b \, , \label{eq:beta}
\ee
the action (\ref{eq:sigma}) becomes invariant under the rigid symmetry group $G$. In formulas this implies that for any such a vector field $v$ on $M$, an infinitesimal change of the fields induced by $\delta X^i := X^* v^i$ leaves the action \eqref{eq:sigma} invariant (up to a boundary term $\int_{\partial \Sigma} X^*\beta$, that is usually considered irrelevant at this point).

More generally, the $B$-contribution in (\ref{eq:sigma}) can be replaced by a Wess-Zumino term
\be S_{W\!Z} = \int\limits_{\tilde{\S}}  \tilde{X}^* H , \label{eq:WZ}\ee
the invariance condition (\ref{eq:alpha}) being generalized to
\be \iota_v H = \rd \a \, . \label{eq:alpha}
\ee
For $H=\rd B$ and  $\a = \b- \iota_v B$ this reproduces the previous situation (the boundary of $\tilde{\S}$ is assumed to be $\Sigma$ and $\tilde{X}$ restricts to $X$ on the boundary by assumption), but the variational problem is well-defined also more generally for $H$ a closed $d+1$-form. While gauging the rigid symmetry group $G$ of (\ref{eq:sigma}) in the case of vanishing $\beta$ is effected by minimal coupling (reviewed below), in the presence of a Wess-Zumino term it is in general even obstructed \cite{Hullbefore, Hull, Stanciu}: Only if $H$ permits an equivariantly closed extension, the sigma model with Wess-Zumino term can be gauged. We will come back to this issue below.

\section{Standard Methods of Gauging and Composite Fields}
The kinetic term in (\ref{eq:sigma}) does not pose any problem in gauging. With (\ref{eq:Kill}) being satisfied, its rigid G-symmetry is lifted to a local one, ${\cal G} \equiv  \mathrm{Maps}(\Sigma,G)$, merely by minimal coupling: This is effected by introducing Lie algebra valued 1-forms $A\equiv A^a e_a \in \Omega^1(\S,\g)$, $e_a$ denoting a basis of $\g$, and replacing $\rd X^i$ everywhere within the functional by the covariant derivative $\rD_A X^i \equiv \rd X^i + v^i_a(X) \, A^a$, where $i=1, \ldots , n$ and $v_a := \rho(e_a)\equiv \rho_a^i(X)\partial_i$ for $a = 1, \ldots,  \dim \g$. Thus,
the gauging of \eqref{eq:standardsigma} is effected by
\begin{equation}
S_{kin}[X^i,A^a] :=  \int\limits_\Sigma  \frac{1}{2} g_{ij}(X) \, \rD_A X^i \wedge * \rD_A X^j \, .
\end{equation}

Let us next focus on the Wess-Zumino term, generalizing the B-contribution of (\ref{eq:sigma}). As mentioned, minimal coupling does not work in this case in general (for example, it would produce a triple $A$ contribution, which would not reduce to a two-dimensional boundary term even after variation). However, as often with Wess-Zumino terms, the general construction can be obtained from the special case $H = \rd B$ upon eliminating all terms that contain $B$ explicitely. To apply minimal coupling for a $B$-term, we in addition need to require $\beta$ to vanish in (\ref{eq:beta}). In this special case we have to identify $\alpha$ of eq.~(\ref{eq:alpha}) with $-\iota_v B$; in fact, so for each $v_a$:
\begin{equation}\label{sim}
\alpha_a \sim - \iota_{v_a} B \, .
\end{equation}
A minimally coupled 2-form $B$ consists of three terms, one that is $B$ itself, which we will rewrite as (\ref{eq:WZ}) using Stokes theorem, one that is linear in $A^a$, and one that it quadratic in these gauge fields:
\begin{equation}\label{min}
\int_\Sigma \frac{1}{2} B_{ij}(X) \rD_A X^i \wedge * \rD_A X^j = \int\limits_{\tilde{\S}}  \tilde{X}^* (\md B)  - \int_\Sigma B_{ij} v_a^i(X) A^a \wedge \md X^j - \frac{1}{2} B_{ij}  v_a^i v_b^jA^a \wedge A^b \, .
\end{equation}
Using the identification \eqref{sim}, the last integrand can be rewritten for example according to $\frac{1}{2} X^*(\iota_{v_b} \iota_{v_a} B) A^a \wedge A^b \sim - \frac{1}{2}v_b^i \alpha_{ai} A^a \wedge A^b$. Introducing composite gauge fields according to the pattern suggested by this rewriting,
\begin{equation}
A_i := \alpha_{ai} (X) A^a \quad , \qquad V^i := v_a^i(X) A^a \, , \label{eq:composite}
\end{equation}
where we stressed that these new gauge fields depend on the old ones as well as the scalar fields $X^i$, the minimally coupled action \eqref{min} can be rewritten as
\begin{equation} \label{eq:top}
S_{top} = \int_{\Sigma} A_i \wedge \rd X^i - \frac{1}{2} A_i \wedge V^i + \int\limits_{\tilde{\S}}  \tilde{X}^* H \,.
\end{equation}
Adding to this the minimally coupled kinetic term, $S_{DSM}:=S_{kin} + S_{top}$, rewritten by means of the composite fields as well, we obtain
\begin{equation}\label{DSM}
S_{DSM} =   \int\limits_\Sigma  \frac{1}{2} g_{ij}(X) \, (\md X^i-V^i) \wedge * (\md X^j-V^j) + \int_{\Sigma} A_i \wedge \rd X^i - \frac{1}{2} A_i \wedge V^i + \int\limits_{\tilde{\S}}  \tilde{X}^* H \, .
\end{equation}
This has precisely the form of the Dirac sigma model introduced in \citep{Kotov-Schaller-Strobl}. In the end, this functional depends only on $H$ and, by means of the identification \eqref{eq:composite},  the couples $(v^a,\alpha_a)$ entering eq.~(\ref{eq:alpha}). It does in particular no more depend explicitely on $B$. It turns out \cite{Hull, Stanciu} that this provides a solution of the gauging problem in two dimensions even in cases where it cannot be obtained from minimal coupling of $B$ for vanishing $\beta$. It gives the general solution whenever a rigid symmetry of $H$ can be gauged without obstructions. It remains to see, next, how the usual formulation of the theory and this alternative one with the composite fields are related to one another in detail, how the gauge transformations map to one another etc. We will now turn to these questions.

\section{Generalized geometry and Lie algebroid morphisms}
First we note that $A_i$ and $V^i$ carry a lower and an upper index and thus naturally live in the cotangent and tangent bundle of $M$ respectively (pulled back to $\S$ by the map $X$, but we will ignore those subtleties for a moment). On the other hand, they are not independent from one another, any choice of couples $(v_a,\alpha_a)$ will single out a subspace $D$ at each point $x \in M$ in which the sum of these two composite fields can take values in. Schematically this is illustrated in Fig.~1, the map $\mu$ being determined by  $(v_a,\alpha_a)_{a=1}^{\dim \g}$:
\begin{figure}[b]
\includegraphics[width=1.\linewidth]{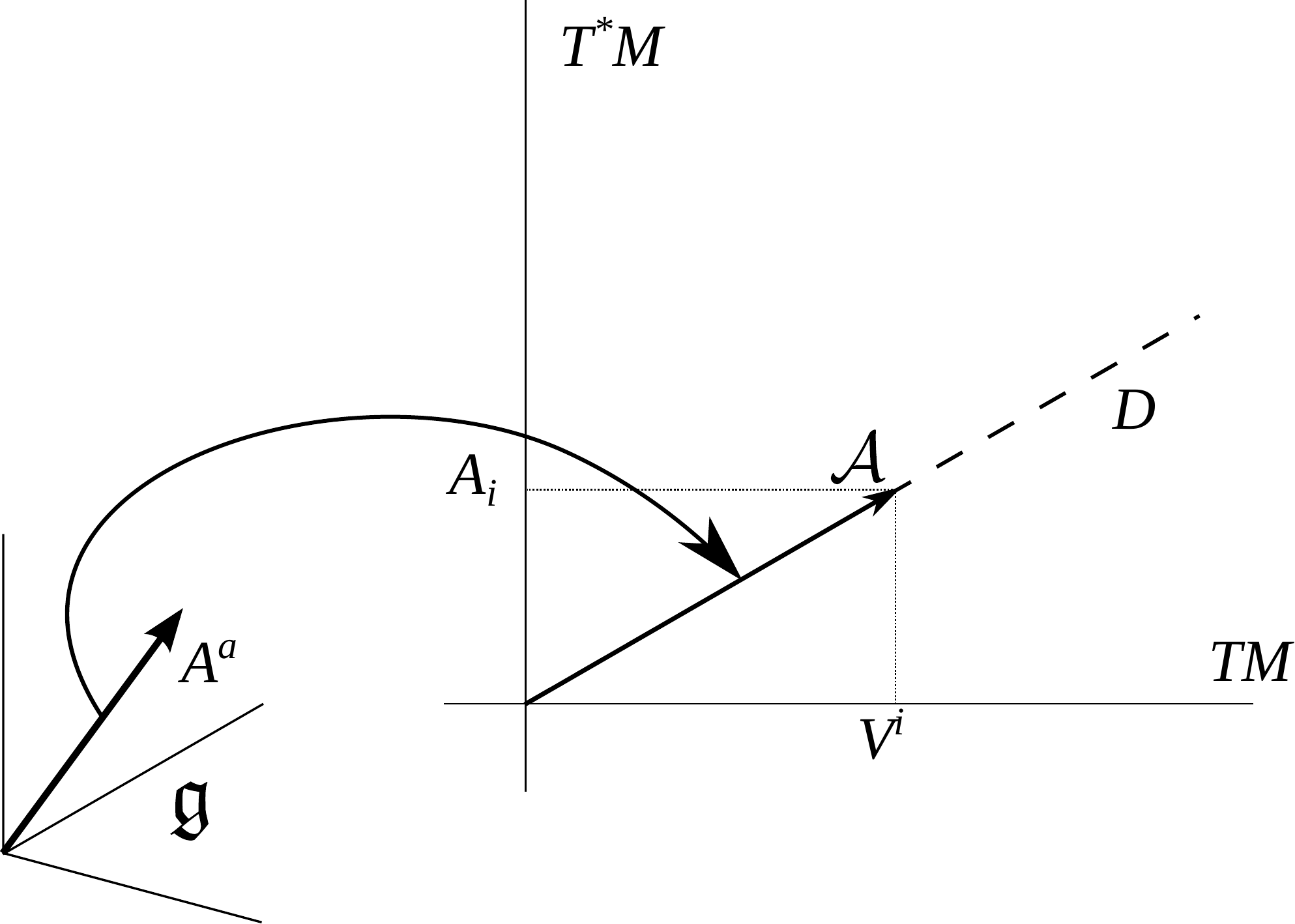}
\caption{\label{fig:composite} At each point $\sigma \in \Sigma$, the standard $\g$-valued gauge fields $A=A^a e_a$ assign to each tangent vector of $\S$ a vector in $\g$; this vector, in turn, is mapped by the morphism $\mu$ to a vector in $D_{X(\sigma)} \subset \left(TM \oplus T^*M\right)|_{X(\sigma)}$.}
\end{figure}
The sum of the tangent and the cotangent bundle $TM\oplus T^*M=:\TTM$ over a manifold $M$ equipped with a closed 3-form $H$ naturally carries what is called a Courant algebroid structure. Any section $\psi$ of $\TTM$ corresponds to a couple of a vector field $v$ and a 1-form $\alpha$. First, $\TTM$ has a natural projection $\rho \colon \TTM \to TM, \psi \mapsto v$. Second, it carries a natural inner product as well as a bracket for the sections:
\begin{eqnarray}
\langle (v,\alpha), (w,\beta) \rangle & = &\iota_v \beta + \iota_w \alpha\, , \label{eq:inner}\\
{}[ (v,\alpha), (w,\beta) ] & = & ([v,w], {\cal L}_v \beta - \iota_w \rd \alpha + \iota_w \iota_v H ) {}\,{} .\label{eq:CD}
\end{eqnarray}
For later purposes we will need some compatibility relation satisfied by the above brackets. The most important property of the Courant-Dorfmann bracket in the above form \eqref{eq:CD} is that it satisfies a left-Leibniz property with respect to itself: for arbitrary sections $\psi_1$, $\psi_2$, $\psi_3 \in \Gamma(\TTM)$ one has
\begin{equation}  \label{Leib}
[\psi_1, [\psi_2,\psi_3]] = [[\psi_1, \psi_2],\psi_3]]+[\psi_2, [\psi_1,\psi_3]] \, .
\end{equation}
On the other hand, the bracket is not symmetric. Instead one finds
\begin{equation}\label{symm}
[\psi,\psi] = \frac{1}{2}(0 ,\md \langle \psi, \psi \rangle) \, ,
\end{equation}
for arbitrary sections in $\TTM$, as one verifies easily from the above definitions. Passing to an anti-symmetrized bracket instead of the above one, the relation \eqref{Leib} is no more satisfied; so with neither of the two brackets one obtains a Lie algebra on the sections  of $\TTM$. One  finally also verifies the compatibility of the bracket with the inner product:
\begin{equation} \label{rhosymm}
\rho(\psi_1) \langle \psi_2, \psi_3 \rangle =  \langle [\psi_1,\psi_2], \psi_3 \rangle +  \langle \psi_2, [\psi_1, \psi_3] \rangle  \, .
\end{equation}

We now intend to understand the map $\mu$ introduced above better. The identification \eqref{sim}, which led us from the
minimally coupled $B$-terms \eqref{min} to the (more general) sigma model (\ref{eq:top}), implies the constraint:
\begin{equation}
\iota_{v_a} \alpha_b + \iota_{v_b} \alpha_a = 0 \, .  \label{eq:iso}
\end{equation}
Moreover, from eq.~(\ref{eq:alpha}) we can induce immediately that for any two vector fields $v$ and $\tilde v$ in the representation of $\g$ one has $\rd {\cal L}_{\tilde v} \alpha_v = \rd \alpha_{[{\tilde v},v]}$. This implies that \emph{up to closed 1-forms} one always has
\begin{equation}
 {\cal L}_{v_a} \alpha_b = C^c_{ab} \alpha_c \, , \label{eq:equiv}
\end{equation}
where $C^c_{ab}$ denote the structure constants of $\g$ in the given basis. In general, there can be obstructions in obtaining the above equation on the nose, even when using the ambiguity in the defninition of $\alpha$s up to closed 1-forms. The situation is very analogous to the one for equivariant moment maps in conventional Hamiltonian mechanics, where Hamiltonians $h_a$ generating a symplectic action $v_a$ are defined only up to constants (closed 0-forms) and in general there may be obstructions for a choice satisfying ${\cal L}_{v_a} h_b = C^c_{ab} h_c$. We will, however, now assume that the above two conditions hold true for some choice of $\alpha$s in (\ref{eq:alpha}). These are also precisely the conditions required for an equivariantly closed extension of $H$ and thus its obstruction-free gauging, cf.~\cite{Hull, Stanciu}.

It is obvious that (\ref{eq:iso}) implies that the subspace $D$ into which $\mu$ maps has to be isotrophic with respect to the inner product (\ref{eq:inner}). Moreover, the rigid invariance condition for $H$, eq.~(\ref{eq:alpha}), implies cancellation of the last two terms in the Courant-Dorfmann bracket (\ref{eq:CD}), and the equivariance condition (\ref{eq:equiv}) is now found to be \emph{equivalent} to the closure of this bracket, yielding $[(v_a,\a_a),(v_b,\a_b)]=C^c_{ab} (v_c,\a_c)$.

By definition, closed, maximally isotropic subbundles of the exact Courant algebroid are called Dirac structures. Dropping the condition of maximality, we call them small Dirac structures. If the rank is permitted to also change from point to point, this gives something that one may call Dirac-Rinehart sheaf. At least in some cases it will be possible to include the image in a honest Dirac structures $D$. So, at least for simplicity of the language, we will now assume this to be the case. We complement this paragraph, however, with a more detailed discussion of this issue in the Appendix A. 

So, at least when understood like explained in the previous paragraph, the image of $\mu$ lies inside a Dirac structure $D$. As mentioned above, the three conditions (\ref{eq:alpha}), (\ref{eq:iso}),  and (\ref{eq:equiv}) are those for the existence of an equivariantly closed extension of the 3-form $H$ \citep{Stanciu}. In \citep{Alekseev-Strobl} these conditions were found to correspond to the existence of a moment map on a Hamiltonian level together with a selection of a Dirac structre. Here we obtain such relation on the Lagrangian level, including the relation of the gauged action functional to the action functional of the Dirac sigma model \citep{Kotov-Schaller-Strobl}, which has precisely the currents of \cite{Alekseev-Strobl} (those appearing in Equation \eqref{currents}) as their constraints, $J_\psi[\varphi]\approx 0$ for any $\psi \in \Gamma(D)$ and any choice of the test functions $\varphi$ (cf.~\cite{Kotov-Schaller-Strobl} for the corresponding details).

Further remarks are in place: The map from the Lie algebra valued gauge fields to the composite ones can have a considerable kernel, in particular if $\dim \g > \mathrm{rk} D = \dim M$ there even has to be a non-trivial kernel. This becomes most transparent if we drop the kinetic term altogether and regard the symmetries of the Wess-Zumino term only. This can easily give an infinite dimensional group. We provide an example: We already remarked above that the condition (\ref{eq:alpha}) together with (\ref{eq:equiv}) is a higher analogue of (pre)symplectic structures and moment maps (cf.~also \cite{Rogers}). Let $(M_0,\omega_0)$ be a $2m$-dimensional ordinary symplectic manifold and consider $M := M_0 \times \R^+$ with the 2-symplectic form $H=\omega_0 \wedge \rd C$ (where $C$ is the strictly positive coordinate on $\R^+$). Now \emph{any} Hamiltonian vector field $v_f$ on $M_0$ lifts canonically to a 2-Hamiltonian vector field on $M$ (we denote it by the same letter), and thus to a rigid symmetry of $H$ in the sense of (\ref{eq:alpha}), where $\alpha_f := C \rd f$ provides a choice of
2-Hamiltonians satisfying both conditions (\ref{eq:iso}) and (\ref{eq:equiv}), as one easily verifies. Here clearly $\g = \mathrm{Ham}(M_0)$ is an infinite dimensional group. In this example, the Dirac structure can be identified with the graph of an $H$-twisted Poisson bivector $\Pi = \frac{1}{C} \Pi_0$, $\Pi_0$ denoting the Poisson bivector of $\omega_0$ (lifted trivially to $M$). Since here always $v_f = \iota_{\a_f} \Pi$, the composite gauge field $V^i$ can be expressed uniquely in terms of the gauge field $A_i$ by means of $V^i = A_j \Pi^{ji}(X)$. The gauged sigma model \eqref{eq:top} then takes the form of the twisted Poisson sigma model \cite{Klimcik-Strobl}
\begin{equation} \label{HPSM}
S_{HPSM} = \int_{\Sigma} A_i \wedge \rd X^i + \frac{1}{2} \Pi^{ij}(X) A_i \wedge A_j + \int\limits_{\tilde{\S}}  \tilde{X}^* H \,,
\end{equation}
which gave rise to the notion of a Wess-Zumino or $H$-twisted Poisson structure and  found subsequently to be a particular Dirac structure
\citep{Severa-Weinstein}.

Thus the functional \eqref{HPSM}  with $\Pi = \frac{1}{C} \Pi_0$ and  $H=\omega_0 \wedge \rd C$ provides a gauging of the WZ-action
\eqref{eq:WZ} w.r.t.~the infinite dimensional gauge Lie algebra  $\g = \mathrm{Ham}(M_0)$. Any standard method, such as the one of an ordinary equivariantly closed extension \cite{Stanciu}, needs for this an infinite number of gauge fields $A^a$. By the present reformulation, which permits to forget about the fact that one may consider $A_i$ as composite fields, one needs only $n=2m+1$ 1-form gauge fields. In \cite{Salnikov-Strobl} this was related to an alternative, non-standard equivariant extension applicable in the context of at least Dirac structures (but with adequate modification potentially also to higher Lie algebroids, following the underlying ideas in \cite{Kotov-Strobl}).

We may use the same example to show that there may be \emph{different} Dirac structures relating  to the same Lie algebra $\g$ and thus different gauged actions. This corresponds to different equivariant extensions of $H$ or, equivalently, different choices of 2-Hamiltonians $\a$. In the above example we could equally well have chosen $\a_f = -f \rd C$, which singles out another $n$-dimensional subbundle $D$ of $\TTM$. It is also a Dirac structure, but one that is for example not the graph of any bivector field. Both the gauge invariant functionals are of the form (\ref{eq:top}), the different details lying in the map $\mu$. Only in the first example it is of the form of a twisted PSM. 

The map $\mu$ can be viewed upon as a Lie algebroid morphism: Any Lie algebra action on a manifold gives rise to a socalled action Lie algebroid structure on  $E=M \times \g$ (cf., e.g., \citep{Weinstein}). Although the Courant algebroid $\TTM = TM \oplus T^*M$ is not a Lie algebroid (but rather a particular Lie 2-algebroid), any Dirac subbundle $D$ is canonically equipped with a Lie algebroid structure.\footnote{This holds also true for small Dirac structures defined in the Appendix. If the image of $\mu$ is not of constant rank and only a Dirac-Rinehart sheaf \emph{and} if in addition it turns out that there is no embedding of the image into a small Dirac structure, then the notion of the Lie algebroid morphism has to be replaced by an algebraic analogue, cf also Appendix A.} In particular, the bracket (\ref{eq:CD}) becomes a Lie bracket upon this restriction, as obvious from combining the Equations \eqref{Leib} and \eqref{symm} with the fact that a Dirac structure $D$ is isotropic, which yields $\langle \psi,\psi \rangle =0$ for any $\psi \in \Gamma(D)$. The map $\mu \colon A \to D, (x,\xi^a e_a) \mapsto (x, \xi^a (v_a^i \partial_i \oplus \alpha_{ai}  \rd x^i))$ is a morphism of Lie algebroids, as follows from the fact that it was seen above to preserve the brackets and that the action of $\xi$ and $\mu(\xi)$ agree on $M$ by definition of the anchor maps.

The existence of a Lie algebroid morphism from the action Lie algebroid $E$ to a Dirac structure $D$ is not yet sufficient for the gauging to exist. We need in addition that the image $(v(\xi),\alpha(\xi))$ of any \emph{constant} section $\xi \in \Gamma(E)$ satisfies the ``rigid symmetry condition''  \eqref{eq:alpha}. Here ``constant'' is well-defined, since $E$ comes as a trivialized bundle, $E=M\times \g$, thus carrying a global tele-parallel structure; such sections are evidently in bijection to elements of the Lie algebra $\g$. It is sufficient to check  \eqref{eq:alpha} on a basis $e_a$ of the Lie algebra, moreover, since this condition is $\R$-linear. However, it is \emph{not} $C^\infty(M)$-linear and thus will not hold true for all sections in $D$. On the Lagrangian level this condition is related to the original rigid symmetry discussed in section 2. On the Hamiltonian as presented in \cite{Alekseev-Strobl} it is more hidden. There, the discussion being independent of action functionals or their corresponding Hamiltonians, it appears in the \emph{existence} of the currents $J_{(v,\alpha)}[\varphi]$: given a $\g$-action on the loop space, one asks it to have a Hamiltonian lift to the cotangent bundle equipped with the canonical symplectic structure twisted by a transgression of the closed 3-form $H$. This is tantamount to the condition  \eqref{eq:alpha} in that framework.

In general the image of $\mu$ does not need to span the whole Dirac structure $D$. If it does, it was shown in \citep{Kotov-Schaller-Strobl} that the model
\begin{equation}
S_{DSM} = \int\limits_\Sigma  \frac{1}{2} g_{ij}(X) \, (\md X^i-V^i) \wedge * (\md X^j-V^j) + S_{top}
\end{equation}
is topological (in the sense that the space of solutions to the Euler Lagrange equations modulo the gauge symmetries is a finite dimensional space). If the image of $\mu$ smears out a subbundle $F \subset D$ only, the model remains ``physical''. A typical example is the G/H WZW-model \citep{Gawedzki}; only in the case that the subgroup $H$ is all of $G$, it becomes topological, otherwise it describes strings moving on a coset space.
In general, moreover, the rank of $\mu$ can change from point to point in $M$ and its image does not define a vector bundle, but only a sheaf of ${\cal O}_M$-modules (cf.~also the discussion in Appendix A). On the other hand, 
let us remark that even at a fixpoint of the $\g$-action, where all $v^a$s vanish, continuity may require some $\alpha_a$s to be nonzero at that point. 
This happens for instance for the adjoint action of a semi-simple Lie group on itself, for which there exists very well a Dirac structure.
Finally, in general, a Dirac structure can happen to be neither the graph of a two-form $B$ nor of a bivector $\Pi$ globally (even after a change of the splitting in the exact Courant algebroid, cf.~\citep{Kotov-Schaller-Strobl}); there is a characteristic class that can be associated to this obstruction, obtained first by A.~Alekseev (cf.~\cite{GeneralizingGeometry} for the construction). 
An example of such a Dirac structure is provided by the G/G WZNW model, with the Dirac structure being the above-mentioned one corresponding to the adjoint action (cf., e.g., \cite{Kotov-Schaller-Strobl}).


\section{Gauge transformations and the Courant bracket}
We now turn to the gauge transformations of the composite fields. Such as the gauge fields $A^ae_a \in \Omega^1(\Sigma,\g)$ are mapped to the composite gauge fields ${\cal A} \in  \Omega^1(\Sigma,X^*D)$ by the map $\mu$ (which also includes and uses the map $X \colon \S \to M$), cf.~Figure 1, likewise this applies also to the gauge parameters $\varepsilon^a e_a \in C^\infty(\S,\g)$, which get mapped to Dirac-structure valued functions. In analogy to (\ref{eq:composite}), this corresponds to
\begin{equation}
\a_i := \varepsilon^a(\sigma)  \alpha_{ai} (X(\sigma)) \quad ,  \qquad v^i :=\varepsilon^a(\sigma) v_a^i(X(\sigma))  \, . \label{eq:compepsilon}
\end{equation}

Now let $(w,\beta) \in \Gamma(\TTM)$ be any couple of a vector field and a 1-form over $M$. Then a direct computation starting from $\delta_\varepsilon A^a = \rd \varepsilon^a + C^a_{bc} A^b \varepsilon^c$ (together with $\delta_\varepsilon X^i = v^i\equiv \varepsilon^a v_a^i$ and the identities satisfied by $v_a$ and $\alpha_a$) yields the remarkable formula
\begin{eqnarray}\label{eq:deltaCourant}
\delta_\varepsilon (w^i A_i + \beta_i V^i) &=&\rd^1 ( \iota_v \beta + \iota_w \alpha) +  \\&&\! \! [v,w]^i A_i
+ ({\cal L}_v \beta - \iota_w \rd \alpha + \iota_w \iota_v H )_i V^i, \nonumber
\end{eqnarray}
to be compared with the two structural equations (\ref{eq:inner}) and (\ref{eq:CD}) of the exact Courant algebroid.
Here $\rd^1$ denotes the de Rham differential on $\Sigma$, but acting only on the explicit $\S$-dependence. Likewise, the derivative terms in the second line are understood as acting only on $M$, reproducing precisely the Courant-Dorfmann bracket. On a more formal level, it is preferable to interpret the formulas as living on the product of $\Sigma$ and $M$. Then $\md^1\equiv \md_\Sigma$ is the deRham differential on the first and $\md\equiv \md_M$ the one on the second factor. Likewise, the bracket of vector fields on $M$ are only well-defined in this setting, while not so in the pull-back bundle over $\Sigma$ for example. Only in a last step then one takes the pullback by $X$ of functions on $M$ to become functions on $\Sigma$. (Certainly this information can be retrieved from knowing formulas like \eqref{eq:deltaCourant} on \emph{all} of $\Sigma \times C^\infty(\Sigma,M)$, the second factor corresponding to the space of maps $X \colon\Sigma \to M$).

We remark that the first line vanishes, if the section $(w,\beta)$ takes values in $D$ (this changes if it also depends explicitly on $\Sigma$, precisely by its $\rd^1$ derivative in that case, cf.~Equation \eqref{deltaAP2} below). Furthermore, the last two terms in the second line cancel out against one another due to the rigid symmetry condition (\ref{eq:alpha}); thus the second line can be rewritten also simply
as $({\cal L}_v w)^iA_i + ({\cal L}_v \beta)_iV^i$. This implies in particular 
\begin{equation} \label{deltaVA}\delta_\varepsilon V^i = \rd^1 v^i + v^i{}_{,j} V^j \; , \quad
\delta_\varepsilon A_i = \rd^1 \alpha_i - v^j{}_{,i} A_j \, .
\end{equation}
Whichever of the above, under the given assumptions equivalent formulas one uses, one always finds that the gauge transformations corresponding to those obtained from the original ones, follow their Lie algebra $(C^\infty(\Sigma,\g),[\cdot, \cdot])$: \begin{equation}\label{epsilons}
[\delta_\varepsilon, \delta_{\bar{\varepsilon}}] = \delta_{[\varepsilon, \bar{\varepsilon}]}
\end{equation} for any choice of  $\varepsilon, \bar{\varepsilon} \in  C^\infty(\Sigma, \g)$. On the other hand, the Poisson sigma model, which is a special case of (\ref{eq:top}), is maybe the simplest prototype of a theory with a so-called ``open gauge algebra'' \cite{Batalin-Vilkovisky,Henneaux-Teitelboim}. Moreover, having observed that $\mu \colon M \times \g \to D$ is a morphism of Lie algebroids, the map (\ref{eq:compepsilon}) should provide also a morphism of the ((closed)) Lie algebra of gauge transformations. How do these obeservations go together?

To address this question properly and in the most elegant way, we first reformulate Equation \eqref{eq:deltaCourant} in a form more adapted to the Courant structure, generalizing it simultaneously to the case where also
the initial section $(w,\beta)$ can depend additionally on $\Sigma$, since the second line in \eqref{eq:deltaCourant} is of this nature anyway. For this purpose we consider $\Phi \equiv (w,\beta)\in C^\infty(\Sigma) \otimes \Gamma(\TTM)$ and call the combination appearing on the left of Equation \eqref{eq:deltaCourant} simply
\begin{equation}\label{APhi}
{\cal A}_\Phi := w^i A_i + \beta_i V^i \equiv \langle \Phi , {\cal A}\rangle \, ,
\end{equation}
where we used the notation of Equation \eqref{eq:inner} for the bracket on the right-hand-side. Depending on the context, one may interpret the expression on the r.h.s.~also as being considered under the pullback by $X \colon \Sigma \to M$ so as to be defined on the field space. We follow more physics-oriented conventions here, not distinguishing these situations too pedantically in favor of a simplified notation.

If we then call the elements $(v,\alpha)$ in the image of the map $\mu$ by
$\Psi \in  C^\infty(\Sigma) \otimes \Gamma(\TTM)$, the formula \eqref{eq:deltaCourant} takes the simple form
\begin{equation}\label{deltaAP2}
\delta_\Psi {\cal A}_\Phi = \langle \rd^1 \Psi , \Phi \rangle + {\cal A}_{[\Psi,\Phi]} \, .
\end{equation}
Here the derivative $\rd^1$ acts only on $\Psi$, which contains the parameters of the gauge transformations, for which reason we wrote this derivative inside the inner product, which, in turn, is not effected by this derivative.

Before continuing to now calculate the commutator of gauge transformations in this language, some further remarks may be in place. First, to understand that $\rd^1 \Phi$ is well-defined without the introduction of a connection on $M$, we may want to interpret it as living on the product $\Sigma \times D$ or $\Sigma \times \TTM$, in which case we can perform the derivative on the function part on the first factor without problems; the pullback by $X$ is then to be taken afterwards only. These ``explicit'' derivatives have a long tradition in physics and always can be reinterpreted in such a way. Second, the symmetries arising from the map $\mu$ by means of \eqref{eq:compepsilon} yield sections $\Psi$ that live inside the Dirac structure $D$ (or a Dirac-Rinehart sheaf ${\cal{D}}$) and not all of $\TTM$. It is tempting, however, to take the transformations \eqref{deltaAP2} as given for any section $\Psi \in C^\infty(M)\otimes \Gamma(\TTM)$ and to subsequently consider the closure of such transformations as a condition on these parameters, in some analogy with
\cite{Alekseev-Strobl}. In fact, to make this analogy more transparent, we have chosen to write the parametrizing sections according to $\Psi = \varphi \psi$ in the Introduction, with $\varphi \in C^\infty(\Sigma)$ playing the role of (arbitrary) test functions and $\psi$ the section in the Courant algebroid to eventually be restricted to a Dirac structure. Certainly, any section $\Psi$ is a sum of such factorized ones and a parametrization like this does not change the picture, since at the end one always passes to the linear span of the symmetry generators to obtain the symmetry algebra.

We now assume $\Psi$ and $\bar{\Psi}$ to be arbitrary such symmetry generators and we will study their commutators when acting on the variables ${\cal A}_\Phi$, where we permit $\Phi$ to be likewise a $\Sigma$-dependent section in $\TTM$ for generality. The calculation will be surprisingly elegant and short upon using the structural identities \eqref{Leib}, \eqref{symm}, and \eqref{rhosymm} of the Courant algebroid. Evidently, for reapplying \eqref{deltaAP2} to itself, we also need to know, how $\Psi$ acts on a ($\Sigma$-dependent) function $f$ on $M$. Clearly this is given by
\begin{equation} \label{deltaX}
\delta_\Psi f = \rho(\Psi) f \, ,
\end{equation}
the application of the vector field part of $\Psi$ to the function $f$ on $M$ (ignoring all the $\Sigma$-dependence, $f$ being permitted to simultaneously be a differential form on $\Sigma$).

Now we are ready to perform the calculation and we will do it in all detail:
\begin{eqnarray}
[\delta_{\Psi}, \delta_{\bar{\Psi}}] \,{\cal A}_\Phi &=&
\delta_\Psi\left(\langle \rd^1 \bar{\Psi} , \Phi \rangle + {\cal A}_{[\bar{\Psi},\Phi]} \right) - \left( \Psi \leftrightarrow \bar{\Psi}\right)  \\
&=& \rho(\Psi) \langle  \rd^1 \bar{\Psi} , \Phi \rangle +
\langle \rd^1\Psi ,[\bar{\Psi},\Phi]\rangle + {\cal A}_{[\Psi,[\bar{\Psi},\Phi]]} - \left( \Psi \leftrightarrow \bar{\Psi}\right)\nonumber\\
&=&  \langle [\Psi, \rd^1 \bar{\Psi}] -  [\bar{\Psi}, \rd^1 {\Psi}] , \Phi \rangle +  {\cal A}_{[\Psi,[\bar{\Psi},\Phi]] -[\bar{\Psi},[{\Psi},\Phi]]  } \, .  \nonumber
\end{eqnarray}
Here the second term in the second line arose from the inhomogeneity of the transformation property \eqref{deltaAP2} of ${\cal A}$. To arrive from the second to the third line, we first applied Equation \eqref{rhosymm} to the first term, noticing that one of the two resulting terms cancels against the second term due to the anti-symmetrization in $\Psi$ and $\bar{\Psi}$.
To find the new parameter acting on $\Phi$, we may apply Equation \eqref{Leib} to the last term in the equation above, which then transforms into ${\cal{A}}_{[[\Psi,\bar{\Psi}],\Phi]}$. Comparison with \eqref{deltaAP2} shows that we need to transform the first two terms into $\langle \rd^1([\Psi,\bar{\Psi}],\Phi \rangle$, which in fact would follow at once, if the Courant-Dorfmann bracket \eqref{eq:CD} were antisymmetric (note that the bracket is defined completely on $M$ and does not see the derivative along $\Sigma$). In general, however, is not antisymmetric and we pick up an additional term according to Equation \eqref{symm}, which yields
\begin{equation}
[\bar{\Psi}, \rd^1 {\Psi}] + [\rd^1 {\Psi},\bar{\Psi}]= \rd \langle \rd^1 {\Psi} , \bar{\Psi} \rangle \, .
\end{equation}
Here we again used a simplified notation, $\rd$ being as above the deRham differential on $M$ only and we interpret the resulting 1-form on $M$ as a section in $\TTM$ by extending it trivial to the $TM$-part, $\rd f \sim (0,\rd f)$. Putting these observations all together, the symmetry laws \eqref{deltaAP2} and \eqref{deltaX} induce the following commutators on the variables \eqref{APhi}:
\begin{equation} \label{comm2}
[\delta_{\Psi}, \delta_{\bar{\Psi}}] \,{\cal A}_\Phi = \delta_{[\Psi,\bar{\Psi}]} \,{\cal A}_\Phi -\big{\langle} \rd \langle \rd^1 {\Psi} , \bar{\Psi} \rangle,\Phi \big{\rangle} \, ,
\end{equation}
where the brackets on the right-hand side are those defined in Equations \eqref{eq:CD} and \eqref{eq:inner} above. Closure of the gauge symmetries parametrized by sections $\Psi$ does require them to take values inside some Dirac structure $D\subset \TTM$, as we anticipated already in the Introduction: from the above Equation \eqref{comm2}, one concludes Equation \eqref{comm} by means of setting $\Psi := \varphi \psi$ and likewise so for the barred variables.

If, on the other hand, we require $\Psi$ and $\bar{\Psi}$ to be elements of
$C^\infty(\Sigma) \otimes \Gamma(D)$ for a given Dirac structure $D \subset \TTM$, then we indeed obtain the closed algebra
\begin{equation}\label{closed}
[\delta_{\Psi}, \delta_{\bar{\Psi}}] \,{\cal A}_\Phi = \delta_{[\Psi,\bar{\Psi}]} \,{\cal A}_\Phi
\end{equation}
without the ``anomalous term'' on the r.h.s.~of \eqref{comm2} and in accordance with Equation \eqref{epsilons}.

We are now left only with understanding how one reobtains the ``open algebra'' from the above closed one in the case of the general Dirac sigma model or its special case like the $H$-twisted Poisson sigma model \eqref{HPSM}. To simplify the discussion, we focus on this latter case, i.e.~on a Dirac structure which is the graph of a bivector $\Pi$ so that each element can be uniquely parametrized by means of $x \in M$ together with an element  $\alpha \in T_x^*M$ according to  $\alpha_i \, (  \Pi^{ij} \partial_j,\rd x^i) \in D_x \subset \TTM_x$. This defines a Dirac structure, iff \cite{Severa-Weinstein} the bivector field $\Pi$ is $H$-twisted Poisson \cite{Klimcik-Strobl} (cf.~also \cite{Park}), thus satisfying, by definition,  \begin{equation}
\frac{1}{2} [ \Pi,\Pi ] = \langle H , \Pi^{\otimes 3} \rangle \, ,
\end{equation}
where the brackets on the r.h.s.~denote contraction here.
This has the advantage that we can focus on the gauge fields $A_i$, the fields $V^i$s being determined by them (and $X \colon \Sigma \to M$) according to $V^i = -\Pi^{ij} A_j$.

To further simplify the discussion, we first consider the ordinary Poisson sigma model (PSM), resulting from \eqref{HPSM} for $H=0$. Then the condition \eqref{eq:alpha} reduces to $\rd \alpha = 0$.\footnote{\label{footBKS} We use this occasion to correct a mistake at the end of \cite{Bojowald-Kotov-Strobl}: For gauge invariance of the PSM in the present picture as well as the one advocated in \cite{Bojowald-Kotov-Strobl}, the condition $\rd \alpha = 0$ is necessary (and not just the weaker one given in Proposition 9 of \cite{Bojowald-Kotov-Strobl}; the mistake in the proof there occured by not writing the pull-back, which does not commute with $\rd_\Sigma$). There is an option to avoid restriction to  $\rd \alpha = 0$ by changing the picture: either by lifting it to the tangent as in \cite{Salnikov-Strobl} or by working with higher jets.} It is easy to see, that the 1-form part (which determines all of the section in $D$) of the bracket \eqref{eq:CD} stays inside the closed forms: in fact, for any $\alpha$, $\beta \in \Omega^1_{cl}(M)$ one has
\begin{equation}\label{ab}
[\alpha,\beta] = \rd \langle \Pi, \alpha \otimes \beta \rangle
\end{equation}
where  the brackets on the r.h.s.~denote simple contraction again: $\langle \Pi, \alpha \otimes \beta \rangle = \iota_\beta \iota_\alpha \Pi$. Here we used that the vector-field part of a section in $D$ with 1-form part $\alpha$ has the form $v = \iota_\alpha \Pi$. So, also in this way, we find a closed algebra (in fact, in agreement with \cite{Bojowald-Kotov-Strobl}).

The known open algebra of the PSM arises when one regards the parameters of the gauge transformations as being sections in the pullback bundle $X^*T^*M$. This corresponds to elements $\alpha = \alpha_i(\sigma,x) \rd x^i$ in the present picture that are independent of $x$ in some particular coordinate system on $M$. Although closed 1-forms produce closed (even exact) 1-forms under the bracket \eqref{ab}, 1-forms with constant coefficients $\alpha_i$ (in some fixed coordinate system) produce likwise 1-forms if and only if the bivector $\Pi$ is a linear function of $x$ in the chosen coordinate system. This corresponds to the case where the PSM reduces to a BF-theory for some Lie algebra, the dual of which is the target Poisson manifold \cite{Schaller-Strobl}.

Let us make this more explicit. Suppose $\alpha = \epsilon_i(\sigma) \rd x^i$ and $\bar{\alpha} = \bar{\epsilon}_i(\sigma) \rd x^i$, such that $\epsilon$ and $\bar{\epsilon}$ \emph{can} be regarded as living in $\Gamma(X^*T^*M)$.\footnote{To distinguish these parameters from those appearing in Equation \eqref{epsilons}, we used another type of epsilon, $\epsilon$ instead of $\varepsilon$.} The sections $\Psi$ and $\bar{\Psi}$ of Equation \eqref{closed}, which is still applicable here (!), can be identified with these $\alpha$ and $\bar{\alpha}$, respectively. We first show that like this one reproduces the usual gauge transformations \cite{Schaller-Strobl,Ikeda} of the PSM,
\begin{equation} \label{PSM}
\delta_\epsilon A_i = \rd_\Sigma \epsilon_i(\sigma) +
\Pi^{jk}{}_{,i} A_j \epsilon_k \; ,
\end{equation}
from \eqref{deltaAP2} as a special case. Indeed, ${\cal A}_\Phi$ reduces to $A_i$ for $\Phi = (0,\rd x^i)$ and, as we know, the Equation \eqref{deltaAP2} reduces to the second equation in \eqref{deltaVA}, where now
$v^j \equiv \Pi^{kj} \alpha_k$ and thus
\begin{equation}
v^j{}_{,i} \equiv \left(\Pi^{kj} \alpha_k\right)_{\!\!,i} = \Pi^{kj}{}_{,i} \epsilon_k(\sigma) \, .
\end{equation}
Since also $\rd^1 \alpha_i \equiv \rd_\Sigma \epsilon_i$, it is evident that
Equation \eqref{PSM} can be obtained from the present formulas. Let us now  turn to the r.h.s.~of Equation \eqref{closed}. For this we need to calculate the bracket between $\Psi$ and $\bar{\Psi}$, which, according to Equation \eqref{ab}, gives the new parameters
\begin{equation}\label{x}
[\epsilon_i(\sigma) \rd x^i,\bar{\epsilon}_j(\sigma) \rd x^j] = \rd \left( \Pi^{ij}\epsilon_i\bar{\epsilon}_j\right) = \Pi^{kl}{}_{,i}(x) \epsilon_k(\sigma)\bar{\epsilon}_l(\sigma) \rd x^i =: \widehat{\epsilon}_i(\sigma,x) \rd x^i \, ,
\end{equation}
where we recall that $\rd \equiv \rd_M$ and where we depicted the coordinate dependences on the r.h.s.~for clarity.

In the ``standard picture'', with parameters taking values in the pullback bundle, the new parameter applicable for the commutator of two gauge transformations of the type \eqref{PSM}, is of the form
\begin{equation}\label{tildep}
\widetilde{\epsilon}_i(\sigma) = \Pi^{kl}{}_{,i}(X(\sigma)) \epsilon_k(\sigma)\bar{\epsilon}_l(\sigma) \equiv \widehat{\epsilon}_i(\sigma,X(\sigma))
\end{equation}
and is viewed as a function of $\sigma$ only (which, in some sense, it is).  Let us now, on the other hand, specialise the gauge transformation \eqref{deltaVA} for the parameter found at the r.h.s.~of Equation \eqref{x} and corresponding to the bracket $[\Psi,\bar{\Psi}]$. First, as anticipated already above, we observe that although the parameters entering the bracket on the l.h.s.~were independent of $x$, the one on the r.h.s.~is not, except for precisely the case that $\Pi(x)$ is linear in $x$.  Thus there is now no good reason that the transformations \eqref{deltaAP2}---or their special case \eqref{deltaVA}---again lead to a gauge transformation of the form \eqref{PSM} (since this was seen to result from $x$-independent parameters) and, in fact, this will not be the case: To see this in detail,
we specialize the gauge transformations on the r.h.s.~of \eqref{deltaVA} to the parameters appearing on the r.h.s.~of Equation \eqref{x}. Here it is useful to use a more mathematical notation, displaying pullbacks by $X \colon \Sigma \to M$ (or its trivial extension to $X \colon \Sigma \to \Sigma \times M$) explicitly.\footnote{It was precisely this notational inaccuracy that led to the mistake mentioned in footnote \ref{footBKS}.} In detail:
\begin{eqnarray}
[\delta_\epsilon,\delta_{\bar{\epsilon}}] A_i &=& X^* \delta_{[\alpha,\bar{\alpha}]} A_i\\
&=& X^* \rd_\Sigma  \widehat{\epsilon}_i(\sigma,x) - X^*\left( \Pi^{kj}(x)  \widehat{\epsilon}_k(\sigma,x)\right)_{\!\!,i} A_j  \nonumber \\
&=& \rd_\Sigma X^*  \widehat{\epsilon}_i(\sigma,x) - X^* \rd_M  \widehat{\epsilon}_i(\sigma,x)+ \nonumber \\
&& + X^*\left( \Pi^{jk}(x) \right)_{\!\!,i} X^*\widehat{\epsilon}_k(\sigma,x) A_j + X^*\Pi^{jk}(x) X^*\widehat{\epsilon}_{k,i}(\sigma,x) A_j \nonumber \\
&=& \rd_\Sigma \widetilde{\epsilon}_i(\sigma) +
\Pi^{jk}{}_{,i}(X(\sigma))  A_j \widetilde{\epsilon}_k(\sigma) - X^*( \widehat{\epsilon}_{i,k}(\sigma,x)) X^*(\rd x^k + \Pi^{kj}A_j) \, .
\nonumber
\end{eqnarray}
Here we used that the \emph{total} deRham differential commutes with the pullback, which gives $\rd_\Sigma X^* = X^* (\rd_\Sigma + \rd_M)$, as well as
the fact that the 1-forms \eqref{x} are no more constant but remain closed so that $\widehat{\epsilon}_{k,i}=\widehat{\epsilon}_{i,k}$. Explicitly, $\widehat{\epsilon}_{i,j}= \Pi^{kl}{}_{,ij}(x) \epsilon_k(\sigma)\bar{\epsilon}_l(\sigma)$, so that together with \eqref{tildep} we finally obtain the known open algebra
\begin{equation}\label{comm3}
[\delta_\epsilon,\delta_{\bar{\epsilon}}] A_i = \delta_{\widetilde{\epsilon}} A_i -  \Pi^{kl}{}_{,ij}(x) \epsilon_k(\sigma)\bar{\epsilon}_l(\sigma) \left(\rd_\Sigma X^j(\sigma) + \Pi^{jm}(X(\sigma)A_m\right) \, .
\end{equation}
The main lesson to learn from this calculation may be that at least in sigma models as the PSM it is not the best idea to view the parameters of infinitesimal gauge transformations to live in a pullback bundle. Instead, it is better to view them as living on the product space of the base and the target manifold.

It is instructive to also consider the $H$-twisted Poisson sigma model \eqref{HPSM} from this perspective. The main difference is that now the condition \eqref{eq:alpha} prohibits one to choose parameters $\alpha$ that are constant in some coordinate system, as it was underlying the above identification leading e.g.~to Equation \eqref{PSM}. Still, Equation \eqref{deltaVA} and all the general formalism remains applicable also in this case.

We will be more brief now and switch back to the physics notation. Using \eqref{eq:alpha}, one may rewrite the second equation of \eqref{deltaVA} according to
\begin{equation}
\delta_\alpha A_i = \rd_{tot} \alpha_i + C^{jk}{}_i \alpha_k A_j - \alpha_{i,j} F^j \, ,
\end{equation}
where $\rd_{tot} \equiv \rd_\Sigma + \rd_M$, $F^i \equiv \rd X^i + \Pi^{ij} A_j$, and $C^{jk}{}_i$ are the structure functions of the Lie algebroid $T^*M \cong D$ induced by the twisted Poisson Dirac structure $D$ (i.e.~induced by the bracket \eqref{eq:CD} by restriction to the graph of $\Pi$):
\begin{equation}\label{dd}
[\rd x^i , \rd x^j ] = \left(\Pi^{ij}{}_{,k} + \Pi^{i l} \Pi^{jm} H_{lmk} \right) \rd x^k =:  C^{ij}{}_k \rd x^k \, .
\end{equation}
The symmetries of the $H$-twisted Poisson sigma model in their conventional form, parametrized by a section $\epsilon \in \Gamma(X^*T^*M)$, has the form
\begin{equation}\label{deltaHPSM}
\delta_\epsilon A_i =
 \rd_\Sigma \epsilon_i +
C^{jk}{}_{i} A_j \epsilon_k + \frac{1}{2}\epsilon_k \Pi^{kl} H_{lij} F^j \, ,
\end{equation}
generalizing the symmetries \eqref{PSM} of the ordinary PSM to non-vanishing WZ-term $H$ and, simultaneously, providing an example of the Lie algebroid based symmetry-considerations in  \cite{Bojowald-Kotov-Strobl}. As discussed in that paper, both type of symmetries, \eqref{PSM} and \eqref{deltaHPSM}, are not $M$-covariant, while they  can be made to be so by adding a term proportional to $F^i$ upon the usage of an auxiliary connection on $T^*M$.
This is in contrast to the symmetries in the form \eqref{deltaAP2}, which are evidently and inherently covariant. They permit to reproduce \eqref{deltaHPSM} upon choosing a coordinate system in which the symmetric part of the derivatives of $\alpha$ vanish, $\alpha_{i,j} + \alpha_{j,i} = 0$, while the antisymmetric part remains determined by means of Equation \eqref{eq:alpha}.\footnote{This is possible always in a region of constant rank of  $\rd \alpha =\iota_v H$. The resulting formulas work always.} Now the point is that while the bracket \eqref{eq:CD} remains stable with respect to the condition \eqref{eq:alpha}, it does not respect the condition of symmetrized derivatives to remain zero. In fact, upon usage of this condition and \eqref{eq:alpha} for $\alpha$ and $\bar{\alpha}$, one finds a notable simplification to occur:\footnote{Note that this condition is certainly not in contradiction to Equation \eqref{dd}, since $\rd x^i$ does not satisfy the condition \eqref{eq:alpha} assumed here. Also, a Dirac structure being a Lie algebroid, for general 1-forms $\alpha$ and $\bar{\alpha}$ one may deduce  $ [\alpha,\bar{\alpha}]=C^{ij}{}_{k} \alpha_i \bar{\alpha}_j \rd x^k \left((\iota_\alpha \Pi)\bar{\alpha}_i -(\alpha \leftrightarrow \bar{\alpha})\right)\rd x^i$ from \eqref{dd}, which, for the particular $\alpha$s assumed here, can be seen to reduce to the simple expression \eqref{aa} below.}
\begin{equation}\label{aa}
[\alpha,\bar{\alpha}] = 
\Pi^{ij}{}_{,k} \alpha_i \bar{\alpha}_j \rd x^k \, .
\end{equation}
Note that despite the $H$-dependence of the structure functions in Equation \eqref{dd}, it drops out from the above combination in the particular frame chosen (the equation is evidently non-covariant). In this way one obtains the commutator of gauge transformations similarly to and in generalization of Equation \eqref{comm3} to take the form:
\begin{equation}
[\delta_\epsilon,\delta_{\bar{\epsilon}}] A_i = \delta_{\widetilde{\epsilon}} A_i -  \left(\Pi^{kl}{}_{,ij} + \frac{1}{2}(\Pi^{km} \Pi^{ln})_{,(i} H_{j)mn} \right) \epsilon_k\bar{\epsilon}_l F^j \, ,
\end{equation}
where, according to Equation \eqref{aa}, $\widetilde{\epsilon}_i=\Pi^{kl}{}_{,i} \epsilon_k \bar{\epsilon}_l$, and the brackets around the indices $i$ and $j$ on the r.h.s.~denote symmetrization.
This equation is not only much more time-intensive to obtain when starting directly from \eqref{deltaHPSM}, again we see that we can consistently avoid an open gauge algebra of symmetries by giving up the idea that the symmetries should be parametrized by sections in a pullback bundle, replacing it with sections on appropriate product spaces.

\newpage

\section{Conclusion and Outlook}
We summarize the findings of this paper best in terms of a picture, figure \ref{fig:morphism} below.
\begin{figure}[ht]
\includegraphics[width=1.\linewidth]{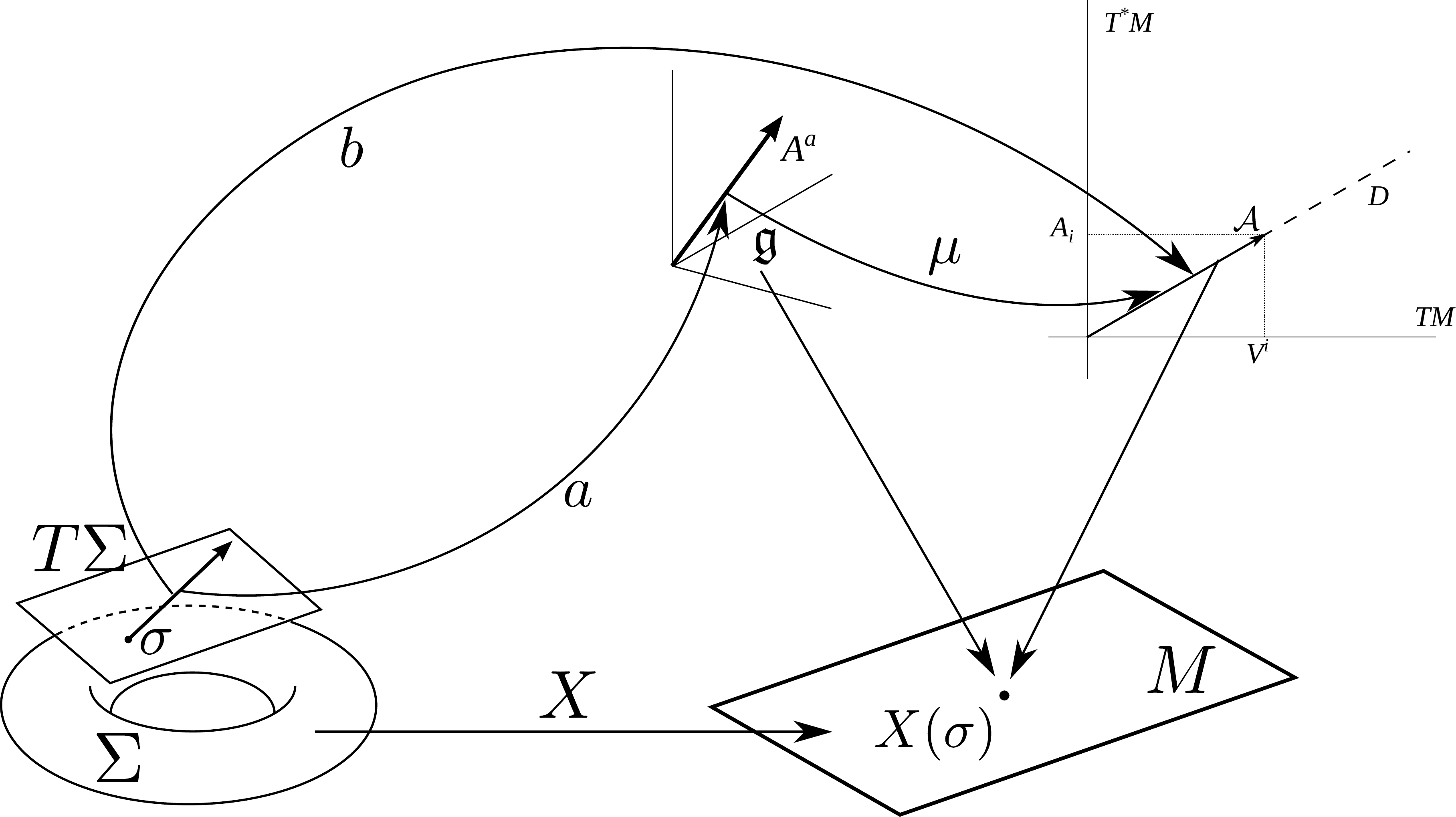}
\caption{\label{fig:morphism} Picture summarizing the findings of the present paper: There is a general sigma model $S_{DSM}$ on a 2d worldsheet $\Sigma$ defined over the vector bundle morphisms $b$ from $T\Sigma$ to the ``generalized tangent bundle'' $\TTM = TM \oplus T^*M$. The conventional gauging of a sigma model with WZ-term is in general obstructed. If it is not, its field content corresponds to a vector bundle morphism $a \colon T\Sigma \to E \equiv M \times \g$ and is equipped with a Lie algebroid morphism $\mu$ from $E$ into a Dirac structure $D\subset \TTM$ such that the gauged model can be obtained as a pullback from $S_{DSM}$, cf.~Equation \eqref{central}.
The Lie algebra based gauge transformation on the l.h.s.~translate into geometrical gauge transformations, based on the Courant-Dorfmann bracket on the r.h.s., cf.~Equation \eqref{deltaAP2}.
 The whole picture generalizes also to non-standard gaugings, as shown elsewhere. Only then the action Lie algebroid $E$ is replaced by a more general Lie algebroid $E \to M$ equipped with a connection. }
\end{figure}
It is known \cite{Hull, Stanciu} that a two-dimensional standard sigma model with target metric $g$ which is invariant under a Lie algebra $\g$, $\CL_{v_a} g = 0$ for all $v_a = \rho(e_a)$, and with a Wess-Zumino term \eqref{eq:WZ} such that the closed 3-form $H$ permits the choice of a $\g^*$-valued 1-form $\alpha=\alpha_a e^a$ satisfying
\begin{equation} \label{rigid}
\iota_{v_a} H = \rd \alpha_a
\end{equation}
as well as the Equations \eqref{eq:iso} and \eqref{eq:equiv}, can be gauged. Given the rigid symmetry conditions \eqref{rigid} to hold true, it is in general not possible to satisfy also these other two equations, even not upon a redefinition of the 1-forms by means of closed additions; gauging of a standard sigma model with WZ-term is thus obstructed in general. And, when gauging is not obstructed, minimal coupling does not give the gauged action functional in general. Let us call this gauged sigma model $S_{gauged}[X^i,A^a]$, or, if we view the map $X \colon \Sigma \to M$ together with the
gauge fields $A^a e_a \in \Omega^1(\Sigma, \g)$ as a vector bundle morphism $a \colon T\Sigma \to M\times \g$, as $S_{gauged}[a]$.

Whenever the obstructions are absent, the Equations  \eqref{eq:iso} and \eqref{eq:equiv} ensure that there is a Lie algebroid morphism $\mu$ from the action Lie algebroid $E = M \times \g$ to a Dirac structure $D$ inside a split exact Courant algebroid $\TTM \equiv TM \oplus T^*M$ with 3-form $H$ (or, more generally, a Lie-Rinehart morphism to ${\cal{D}}$, cf.~Appendix A). Moreover, the gauged sigma model can be obtained from a universal action functional, the Dirac sigma model \eqref{DSM}, which is completely independent of any chosen Lie algebra $\g$ and, as written, even independent of the chosen Dirac structure $D$. Let us make this more explicit: The functional of the DSM as written in \eqref{DSM} can be viewed as a functional on the vector bundle morphisms $b \colon T\Sigma \to \TTM$; we can thus call it $S_{DSM}[b]$.

We remark in parenthesis that the functional \eqref{DSM} was considered in \cite{Kotov-Schaller-Strobl} in the case that $A$ and $V$ cannot be chosen independently, but have to combine into ${\cal{A}} = V \oplus A$ lying in a Dirac structure $D$. In the present perspective, we treat them first as independent. It is only the map $\mu \colon M \times \g \to D\subset \TTM$ that now selects the Dirac structure (or at least a Dirac-Rinehart sheaf).

All the details about the Lie algebra $\g$ and the choice of $\alpha_a$s satisfying the necessary compatibility conditions are now contained in the map $\mu$, defined by $\mu(e_a) = (v_a,\alpha_a)$. The functional  $S_{DSM}[b]$ is independent of all this. The restriction of this functional to the image $\mathrm{im}(\mu)\subset D$ yields the gauging in this picture. In other words,
\begin{equation}\label{central}
S_{gauged}[a] = S_{DSM}[\mu \circ a] \, .
\end{equation}
In other words, $S_{gauged}$ can be viewed as a pullback of $S_{DSM}$; more
precisely, $\mu \colon E \to \TTM$ induces a map $\widetilde{\mu}$ from the 
map space $\{ a \colon T\Sigma \to E\}$ to  $\{ b \colon T\Sigma \to \TTM\}$.  
Then simply $S_{gauged} = \widetilde{\mu}^* S_{DSM}$.

As observed rather recently only \cite{PRL2014}, there are also situations where one can pass to a
gauged version of a given theory even in the absence of an initial symmetry.
In the case of a standard sigma model, one does not need vector fields satisfying an Equation of the form \eqref{eq:Kill}; instead, extra terms of a particular form are permitted on the r.h.s. of \eqref{eq:Kill}. In general then the gauge fields do not need to be Lie algebra valued, moreover, while still they combine with the $X$-fields into a vector bundle morphism such as $a$ in figure \ref{fig:morphism}; only instead of an action Lie algebroid $E=M \times \g$, there will be a generic Lie algebroid $E\to M$. The above picture then still remains correct also in this much more general context, as will be shown in a separate work. Note also that in this context an equation such as Equation \eqref{rigid} does not make sense anymore, but needs to be amended by a connection 1-form on $E$ for the derivative on the r.h.s. This is in fact the same connection that appears already in \cite{PRL2014}.
The analysis then will also provide an explanation for the more elaborate general gauge symmetries of the Dirac sigma model as provided in \cite{Kotov-Schaller-Strobl}. Other evident further questions comprise the generalization of the present picture to higher dimensions and higher algebroids. Also these subjects shall be addressed elsewhere.

Last but not least, there exists a notable relation of generalized geometry to two-dimensional sigma models, tying generalized \emph{complex} geometry to \emph{super-symmetric} 2d sigma models (cf., e.g., \cite{Generalizedbefore,Zabzine}). It will be very interesting to extend the present considerations, and eventually also its above mentioned generalizations, to this setting, too. 

\vspace{-.3cm}
\acknowledgments
T.S.~is grateful to A.~Alekseev and J.~Manschot for discussions.


\appendix
\section{Dirac-Rinehart sheaves and extension to Dirac structures}
In the main text of this paper, we focused on providing the main picture that for any gauged standard sigma model with Wess-Zumino term in two spacetime dimensions there is a Lie algebroid morphism from the action Lie algebroid $M \times \g$ to the Dirac structure $D \subset \TTM$, cf.~Figure \ref{fig:morphism} above. Still, there are some mathematical subtleties that we want to address in this Appendix, refining the observation from a more mathematical perspective, to be taken up in a more profound analysis elsewhere.

First of all, what we really showed only is that by means of the map $\mu$ corresponding to \eqref{eq:composite} we obtain sections $\psi_a \in \Gamma(\TTM)$ that are mutually involutive and isotropic, w.r.t.~\eqref{eq:CD} and \eqref{eq:inner}, respectively. We can take the $C^\infty(M)$-completion of these sections, which, pointwise, corresponds to taking the vector space generated by $\psi_a(x)$ inside $\TTM_x$. 
However, it is not clear that this can be always extended into a full Dirac structure. There may be obstructions for such an extension to exist. 

To clarify the situation a bit further, we first provide some definitions.
\\

{\bf Definition 1:} We call an involutive, isotropic subbundle $D$ of $\TTM$ a \emph{small Dirac structure}. If this bundle has maximal rank, i.e.~$\mathrm{rk} D = \dim M$, we call $D$ a \emph{full Dirac structure} (or simply a \emph{Dirac structure}). A small/full Dirac structure is \emph{regular}, if the (integrable) distribution $F$ induced by the anchor map $\rho \colon \TTM \to TM$ has constant rank.
\\

{\bf Definition 2:} We call a sheaf  $\cal{D}$ of $C^\infty$-submodules of sections of $\TTM$ which is stable under the bracket \eqref{eq:CD} and isotropic with respect to \eqref{eq:inner} a \emph{small Dirac-Rinehart sheaf}. If, in addition,  $\cal{D}$ is almost everywhere of maximal rank, it is called a \emph{(full) Dirac-Rinehart sheaf}, i.e.~for almost each point $x\in M$ there exists a neighborhood $U\ni x$ such that $\mathcal{D}(U)=\Gamma(D_U)$ where $D_U$ is a Dirac structure in $\mathbb{T}U$ (and on overlaps these local Dirac structures glue together consistently). 
\\

Clearly, any small/full Dirac structure defines a small/full Dirac-Reinhart sheaf, but not vice versa, in general. The algebraic notion in the second definition is on the one hand more flexible, on the other hand less suggestive/geometric than the notion in the first definition. A Dirac-Rinehart structure is an example of a Lie-Rinehart pair \cite{Rinehart} in the same way as a Dirac structure is a particular Lie algebroid. 

The map $\mu$ mentioned above is thus in fact a Lie-Rinehart morphism from the Lie-Rinehart pair corresponding to the action Lie algebroid $M \times 
\g$ to a small Dirac-Rinehart sheaf $\cal{D}$. The non-trivial mathematical question that one then can pose is if there \emph{exists} a Dirac structure 
$D$ such that ${\cal{D}}\subset \Gamma(D)$. While we do not know a general answer to this question, in the simplest case we do:
\\

{\bf Proposition:} Any small Dirac structure $
D_s$ can be canonically extended into a full Dirac Rinehart sheaf. If $D_s$ is regular, moreover, it can be canonically extended into a full Dirac structure $D$. 
\\

{\bf Proof:} We first consider the case of a regular small Dirac structure $D_s$.  It comes with the following exact sequence of vector bundles, induced by the anchor map $\rho$ (restricted from $\TTM$ to $D_s$):
$$
  0\to \ker\rho \to D_s \to F\to 0\,.
$$
As a consequence of the isotropy of $D_s$ one also has $\ker\rho\subset \mathrm{ann}(F)\subset T^*M$, where $\mathrm{ann} (F)$ is the annihilator of $F$. Now we define $D:=\mathrm{ann}(F)+D_s$.
By construction, the subbundle $D$ has the same rank as $TM$, and it is obviously isotropic. Moreover, one can easily show that $D$ is also closed 
under the Courant bracket. Indeed, the Courant bracket of two sections of $\mathrm{ann}(F)$ clearly vanishes, while the bracket of
a section $(v^1, \alpha_1)$ of $D_s$ with the section $(0, \beta)$ of $\mathrm{ann}(F)$
gives $[(v^1, \alpha_1), (0, \beta)] = (0, {\cal L}_{v_1}\beta)$.
Since for any $v_2 \in \mathrm{Im}(\rho)$ one has
$\iota_{v_2} {\cal L}_{v_1} \beta = [\iota_{v_2}, {\cal L}_{v_1}] \beta = \iota_{[v_2, v_1]} \beta = 0 $, the section
$(0, {\cal L}_{v_1}\beta)$ is again in $\mathrm{ann}(F)$ (thus $\mathrm{ann}(F)$ is an ideal in $D$ and $D$ 
is closed under the Courant bracket). On the other hand, $D_s\subset D$, so this provides a (canonical) extension of the small Dirac structure $D_s$ to the full Dirac structure $D$.

Given that for \emph{any} small Dirac structure $D_s$ the corresponding distribution $F= \rho(D_s)$ is almost everywhere of constant rank, it is obvious that each $D_s$ can be canonically extended to a Dirac-Rinehart structure by adding to the space of local sections of $D_s$ all local $1-$forms vanishing on $F$. This completes the proof. \hfill $\square$

 \end{document}